\documentclass[journal=jacsat,manuscript=article]{achemso}
\usepackage{graphicx}
\usepackage{multirow}
\usepackage{amsmath}
\usepackage{color}
\usepackage{float}
\usepackage{soul}
\usepackage{extarrows}
\usepackage{hyperref}
\usepackage{setspace}
\usepackage{slashed}
\usepackage{xcolor}
\usepackage{amssymb}
\usepackage{bm}
\usepackage{bbm}
\usepackage{titlesec}
\usepackage{caption}

\usepackage[labelfont=bf,justification=justified]{caption}

\newcommand{\beginsupplement}{%
        \setcounter{table}{0}
        \renewcommand{\thetable}{S\arabic{table}}%
        \setcounter{figure}{0}
        \renewcommand{\thefigure}{S\arabic{figure}}%
     }

\title{Influence of atomic relaxations on the moir\'{e} flat band wavefunctions in antiparallel twisted bilayer WS$_{\text{2}}$ }

\author{Laurent Molino*} \affiliation{ Department of Physics, University of Ottawa, Ottawa, Canada}

\author{Leena Aggarwal*} \affiliation{ Department of Physics, University of Ottawa, Ottawa, Canada}
\author{Indrajit Maity} \affiliation{ Department of Materials, Imperial College London, and the Thomas Young Centre for Theory and Simulation of Materials, London, UK }
\author{Ryan Plumadore} \affiliation{ Department of Physics, University of Ottawa, Ottawa, Canada}
\author{Johannes Lischner} \affiliation{ Department of Materials, Imperial College London, and the Thomas Young Centre for Theory and Simulation of Materials, London, UK }
\author{Adina Luican-Mayer} \affiliation{ Department of Physics, University of Ottawa, Ottawa, Canada}

\email{Luican-Mayer@uottawa.ca}

\date{\today}
\begin{document}
*These authors contributed equally
\begin{abstract}

\textbf{Twisting bilayers of transition metal dichalcogenides (TMDs) gives rise to a periodic moir\'{e} potential resulting in flat  electronic bands with localized wavefunctions and enhanced correlation effects. In this work, scanning tunneling microscopy is used to image a WS$_{2}$ bilayer twisted approximately $3^{\circ}$ off the antiparallel alignment. Scanning tunneling spectroscopy reveals the presence of localized electronic states in the vicinity of the valence band onset. In particular, the onset of the valence band is observed to occur first in regions with a Bernal stacking in which S atoms are located on top of each other. In contrast, density-functional theory calculations on twisted bilayers which have been relaxed in vacuum predict the highest lying flat valence band to be localized in regions of AA' stacking. However, agreement with the experiment is recovered when the calculations are carried out on bilayers in which the atomic displacements from the unrelaxed positions have been reduced reflecting the influence of the substrate and finite temperature. This demonstrates the delicate interplay of atomic relaxations and the electronic structure of twisted bilayer materials. }
\end{abstract}

When a pair of two-dimensional (2D) materials are stacked vertically and then rotated relative to each other, a moir\'{e} superlattice is created. Electrons in such a superlattice experience a moir\'{e} potential which favors electron localization in specific stacking regions, reduces the electrons' kinetic energy, and results in the formation of flat bands with enhanced electron-electron interactions, promoting phenomena such as superconductivity, Mott insulating phases, topological phases, coupled spin-valley states, and interlayer excitons.\cite{shimazaki2020strongly,yankowitz2019tuning,an2020interaction,brotons2020spin,choi2020moire,tran2019evidence,das2020highly}

In contrast to graphene bilayers which exhibit flat bands at a specific ``magic" twist angle, bilayers of transition metal dichalcogenides (TMDs) offer a much larger range of twist angles at which flat bands are realized\cite{PhysRevB.102.075413,wang2020correlated,li2021lattice,li2021imaging, Vitale_2021,waters2020flat,Maity2021reconstruction}, opening broader possibilities for creating previously inaccessible correlated systems. Both TMD homobilayers (in which two layers of the same material are combined) and heterobilayers (in which two different TMD monolayers are combined) have been studied. In heterobilayers, the length scale of the moir\'{e} pattern is less sensitive to the twist-angle than in homobilayers, leading to reduced twist angle disorder in large samples. As a result, to date, several heterobilayer TMDs such as WSe$_{2}$/WS$_{2}$ \cite{li2021imaging,li2021}, MoS$_{2}$/WS$_{2}$ \cite{kobayashi2016modulation,2016shi,zhang2018atomically,hill2016band}, MoS$_{2}$/WSe$_{2}$\cite{pan2018quantum,waters2020flat} and WSe$_{2}$/MoSe$_{2}$ \cite{shabani2021deep} have been investigated experimentally, with reports of correlated phenomena such as superconductivity, Mott insulating phases, generalized Wigner crystals, etc.\cite{regan2020mott,wang2020correlated,tang2020simulation,li2021imaging}. Moir\'{e} patterns of TMD homobilayers are highly sensitive to the twist angle, and thus expected to exhibit more twist-angle disorder. As a consequence, these systems are less studied despite having great potential for inducing strong moir\'{e} potentials \cite{PhysRevB.102.075413,PhysRevLett.121.266401,an2020interaction,van2022quantitative,lin2019determining}. For example, recent scanning probe microscopy experiments on a WSe$_{2}$/WSe$_{2}$ homobilayer have demonstrated twist-angle-dependent lattice reconstructions near a $60^{\circ}$ twist angle resulting in multiple energy-separated ultra-flat valence bands for long moir\'{e} wavelengths (6 nm-12 nm).\cite{li2021lattice} 

 Semiconducting TMDs have metal and chalcogen atoms occupying the lattice sites in their unit cell breaking sub-lattice symmetry. For this reason, twisted TMD bilayers fall into two classes: parallel (P) aligned bilayers with a zero twist angle close to $0^{\circ}$, and antiparallel (AP) aligned bilayers with twist angles near $60^{\circ}$. These two classes exhibit different stacking configurations in the moir\'{e} unit cell which results in different atomic relaxations in the form of both out-of-plane buckling and in-plane reconstructions. In-plane reconstructions can be understood as the result of the layers deforming to increase the area of the energetically preferred stacking order. \cite{weston2020atomic, PhysRevLett.121.266401, Falko2020, PhysRevB.102.075413} For very small twist angles (or twist angles very close to $60^{\circ}$), this reconstruction produces networks of uniform stacking order domains separated by highly strained domain walls. In the case of P-aligned TMD homobilayers, the system tends towards a network of triangular stacking order domains. Due to the lack of inversion symmetry, these domains host interlayer charge transfer leading to an out-of-plane ferroelectric dipole moment.\cite{PhysRevLett.121.266401,shabani2021deep,weston2020atomic,molino2022ferroelectric}
In the case of AP-aligned TMD homobilayers, no such charge transfer occurs. While AP systems do form domain wall networks at angles very close to $60^{\circ}$, they generally exhibit less pronounced in-plane reconstructions than P-aligned bilayers at a given twist-angle offset, leading to a more smoothly varying periodic lattice distortion over the moir\'{e} unit cell. \cite{Bediako2022} In all cases, both out-of-plane buckling and in-plane reconstruction have a profound impact on the electronic properties of TMD homobilayers. In-plane and out-of-plane deformations cause the simple band-folding description of a moir\'{e} pattern band-structure to break down, for example, causing a modification in the real-space position of the valence band edge wavefunctions. Strain is furthermore critical to wavefunction localization responsible for strong correlated effects observed in these systems. \cite{PhysRevB.102.075413, PhysRevLett.121.266401}

As the field of moir\'{e}-induced flat band physics progresses, it is important to explore a larger range of homobilayer systems, especially using scanning probe techniques that can directly reveal the local twist angle. Here, we report room temperature scanning tunnelling microscopy/spectroscopy (STM/STS) studies of moir\'{e} superlattices of $\approx$ 6 nm wavelength in the WS$_{2}$/WS$_{2}$ bilayer system near $60^{\circ}$ twist angle or antiparallel (AP) alignment,  together with extensive large-scale density-functional theory calculations. Importantly, we demonstrate that the localization of the flat-band wavefunctions depends sensitively on the degree of lattice relaxations.

\begin{figure}
        \centering
	\includegraphics[width=1.0\columnwidth]{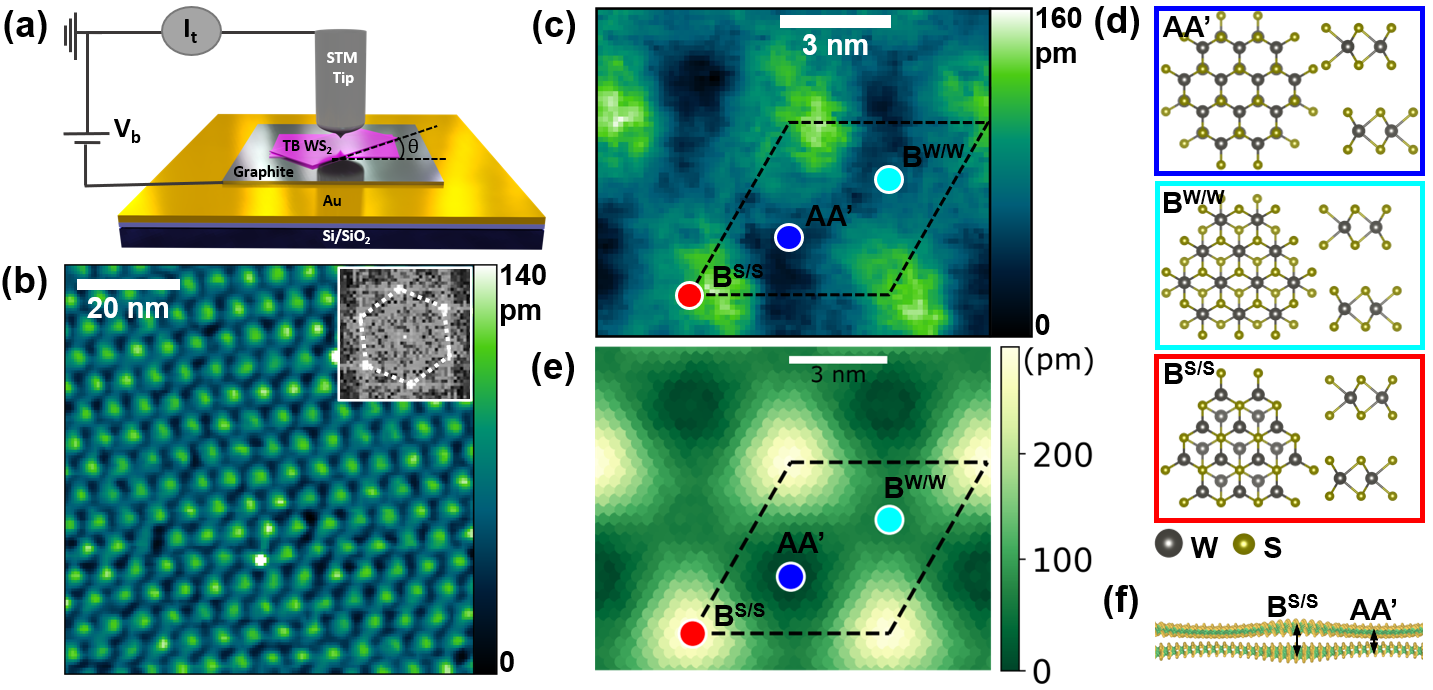}
    \caption{ (a) Schematic of the STM experiment: a twisted bilayer WS$_{2}$ with $\theta \approx 60^{\circ}$ is prepared on a graphite/Au substrate. (b) STM topograph showing a region with  6.0 nm ± 0.2 nm moir\'{e} pattern, corresponding to a twist angle of $57.1^{\circ} \pm 0.1^{\circ}$ (scanning parameters: bias voltage $V_{b} = -1.3 V$, tunneling current $I_{t} = 100$ pA). Inset: Fourier transform of the moir\'{e} pattern showing hexagonal symmetry. (c) STM topograph of 6.0 nm ± 0.2 nm moir\'{e} pattern (scanning parameters: $V_{b} = -1.3 $~V, $I_{t} = 150~$pA). The dashed rhombus corresponds to the moir\'{e} unit cell, with high-symmetry stacking regions ($\mathrm{B^{S/S}}$, $\mathrm{AA}$', $\mathrm{B^{W/W}}$) indicated by circles.  (d) Atomic structure of the three high-symmetry stackings present in an antiparallel aligned twisted WS$_{2}$ homobilayer. (e) Simulated STM topography for a $56.86^\circ$ twisted AP-WS$_2$ homobilayer obtained from first-principles density-functional theory calculations on a relaxed atomic structure. (f) Side view of twisted bilayer WS$_{2}$ showing out-of-plane buckling, exaggerated by a factor of ten for clarity. Locations of maximum and minimum interlayer spacing are indicated. The maximum interlayer spacing, corresponding to B$^{\text{S/S}}$ stacking, is 8\% larger than the minimum interlayer spacing, corresponding to AA' stacking.}
	\label{f1}
\end{figure}

We fabricated the AP-WS$_2$ twisted bilayer using a variation of the ``tear-and-stack" dry transfer method (see Methods section) and performed room temperature STM/STS experiments, as indicated schematically in Figure 1(a). Across the entire sample, we identified moir\'{e} patterns with a range of periods (6 nm-14 nm) (Supplementary Figure S1), corresponding to a twist-angle range of $57^{\circ}$-$58.8^{\circ}$, indicating the presence of twist-angle disorder within the AP-WS$_2$ sample. A large area of uniform hexagonal moir\'{e} pattern with a period of 6.0 nm $\pm$ 0.2 nm (corresponding to a twist angle of $57.1^{\circ} \pm 0.1^{\circ}$) is shown in Figure 1(b). A zoomed-in scanning tunnelling micrograph of this pattern in Figure 1(c) highlights the unit cell and the different high-symmetry stacking regions within it ($\mathrm{B^{S/S}}$, $\mathrm{AA}$', $\mathrm{B^{W/W}}$), with corresponding atomic configurations described in Figure 1(d).

A simulated STM topograph of a $56.86^\circ$ AP-twisted WS$_2$ bilayer obtained from density functional theory (DFT) calculations based on the Tersoff-Hamann approximation~\cite{Tersoff1985theory} is shown in Figure 1(e). Before the DFT calculations, the relaxed atomic positions of the twisted bilayer, shown in Fig. 1(f), were obtained using classical force fields (see Methods). The simulated topograph is in good agreement with the experimental micrograph of Figure 1(c). In particular, the bright regions correspond to $\mathrm{B^{S/S}}$ stacking. This stacking exhibits the largest buckling and therefore also the largest tunnelling matrix element.

\begin{figure}
        \centering
	\includegraphics[width=1.0\columnwidth]{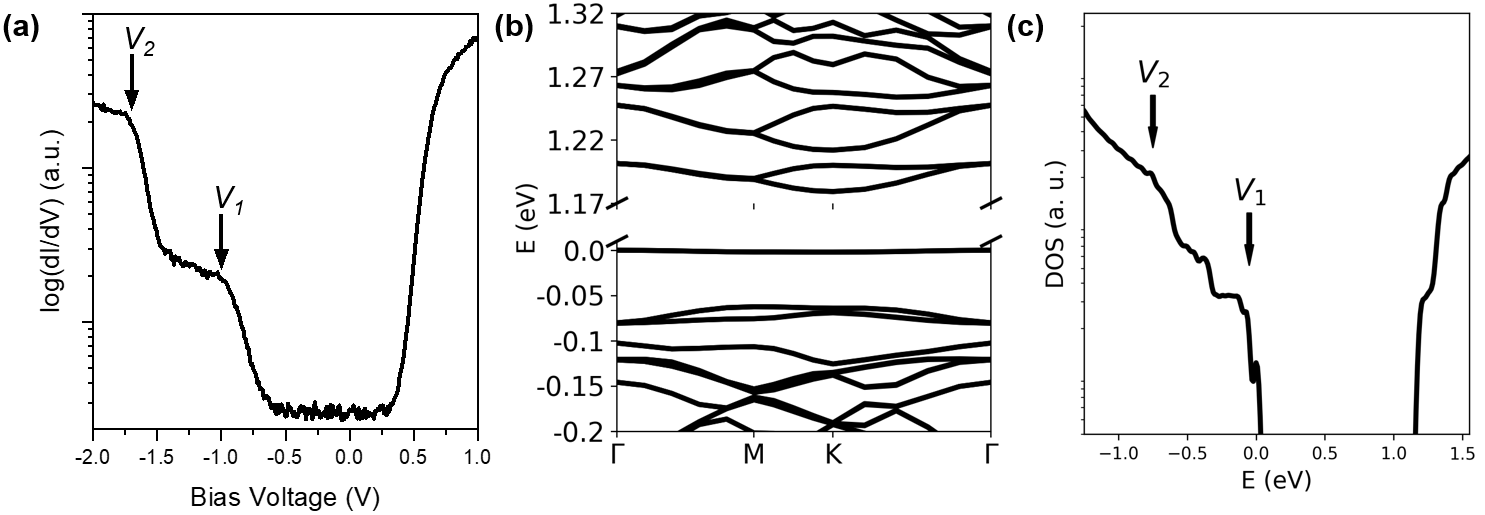}
    \caption{ (a) Average of scanning tunneling spectra acquired over the moir\'{e} region indicated in Supplementary Figure S3.(b) Electronic band structure of $56.86^\circ$ twisted bilayer WS$_{2}$ including atomic relaxations. The valence band maximum is set to 0. (c) Calculated density of states for the same relaxed twisted bilayer WS$_{2}$ as in (b).}
	\label{f1}
\end{figure}

To explore the local electronic properties of the moir\'{e} superlattice in Figure 1(b), we performed scanning tunnelling spectroscopy (STS) measurements. In Figure 2(a) we present the average tunnelling conductance (dI/dV) acquired over five moir\'{e} periods (Supplementary Figures   S2 and  S3). Two notable features are present in the spectrum: a first one (denoted V$_1$) near the valence band edge (VBE), with an energy of $\approx -1.0$ eV, and a second feature V$_2$ deeper in the valence band at $\approx -1.7$~eV.
 
These two features are also found in the density of states (DOS) calculated using DFT for the relaxed atomic structure, see Fig. 2(c). The energy separation between the two features is found to be 0.7 eV, in excellent agreement with the experimental result. We note that the valence band edge of the calculated DOS is much sharper than in the experiment, which is likely a consequence of thermal broadening. 

To understand the origin of the observed features close to the valence band edge, we also present the DFT band structure of the twisted bilayer in Fig. 2(b). This reveals that the highest valence band is extremely flat with a bandwidth of 2 meV. Approximately 60 meV below the highest valence band we find another pair of relatively flat bands which contribute to the sharp feature $V_1$. We also identify another sharp feature in the DOS 0.7 eV below the $V_{1}$, in good agreement with the experiment.

\begin{figure}
        \centering
	\includegraphics[width=1.0\columnwidth]{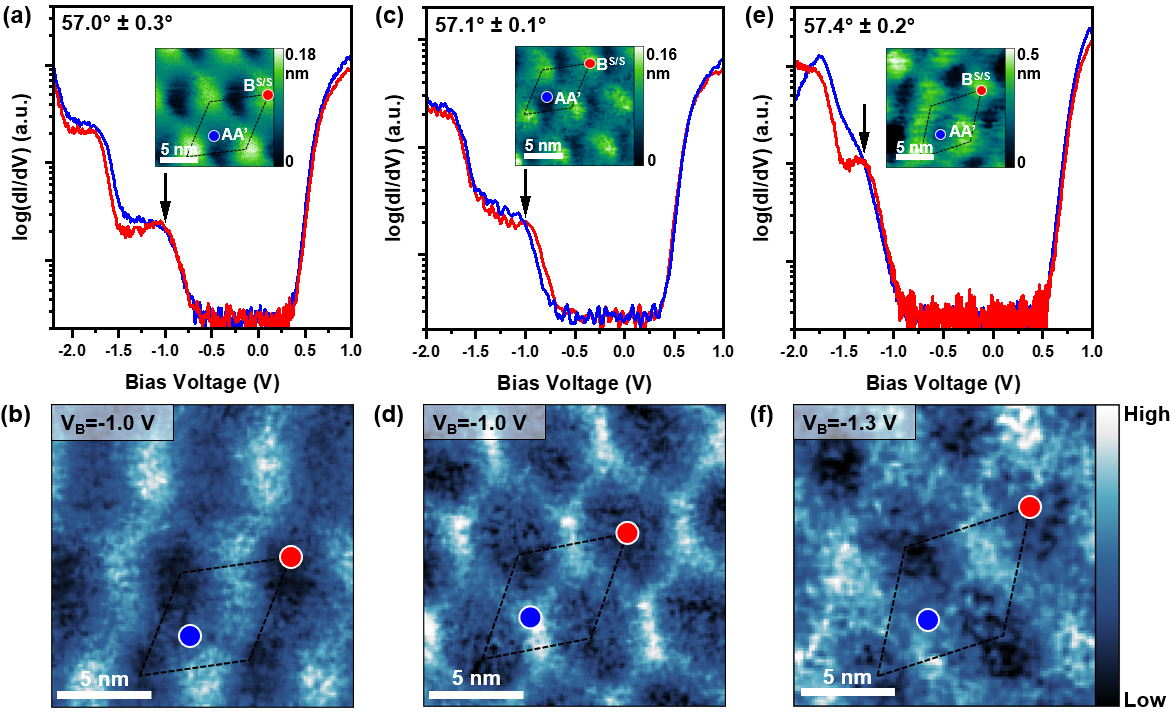}
    \caption{(a),(c),(e) Scanning tunneling spectra acquired at high symmetry sites in three areas having different twist angles as indicated. Inset: Topographic maps indicating the points where the data were taken. (b),(d),(f) dI/dV maps acquired near the valence band edge at bias voltages indicated (tunneling current $I_t = 100 pA$). The high symmetry sites and the moir\'{e} unit cell are indicated in each map.}
	\label{f1}
\end{figure}

Taking advantage of the spatial resolution of a STM/STS experiment, we measured dI/dV maps corresponding to the feature V$_1$. In Figure 3 we present scanning tunnelling spectra collected at AA' and $\mathrm{B^{S/S}}$ stacking regions taken on three areas with slightly different twist angles: $57.0^\circ\pm 0.3^\circ$ [Figure 3(a),(b)], $57.1^\circ\pm 0.1^\circ$ [Figure 3(c),(d)] and $57.4^\circ \pm 0.2^\circ$ [Figure 3(e),(f)]. The insets show the associated topographic image including the high-symmetry points and unit cell. These maps demonstrate that the feature V$_1$ is associated with localized states that form a hexagonal pattern with low intensity on the $\mathrm{B^{S/S}}$ sites and high intensity on the AA' sites. The spectra taken at the AA' and $\mathrm{B^{S/S}}$ sites show that the valence band onset occurs at the $\mathrm{B^{S/S}}$ site, suggesting that the highest-lying valence band is localized at this site (indicating that the localized state associated with V$_1$ is \emph{not} the highest lying valence band). While this difference between the two sites is clearly visible for the $57.1^\circ \pm 0.1^\circ$ system, the separation between the two spectra is much smaller for the other two systems (for the $57.0^\circ \pm 0.3^\circ$ system the onsets at the AA' and the $\mathrm{B^{S/S}}$ sites occur almost at the same energy). We also note that more significant differences between the various twist-angle regions are found further away from the valence band onset indicating that the electronic structure is highly sensitive to the local twist angle. 
\begin{figure}
        \centering
	\includegraphics[width=1.0\columnwidth]{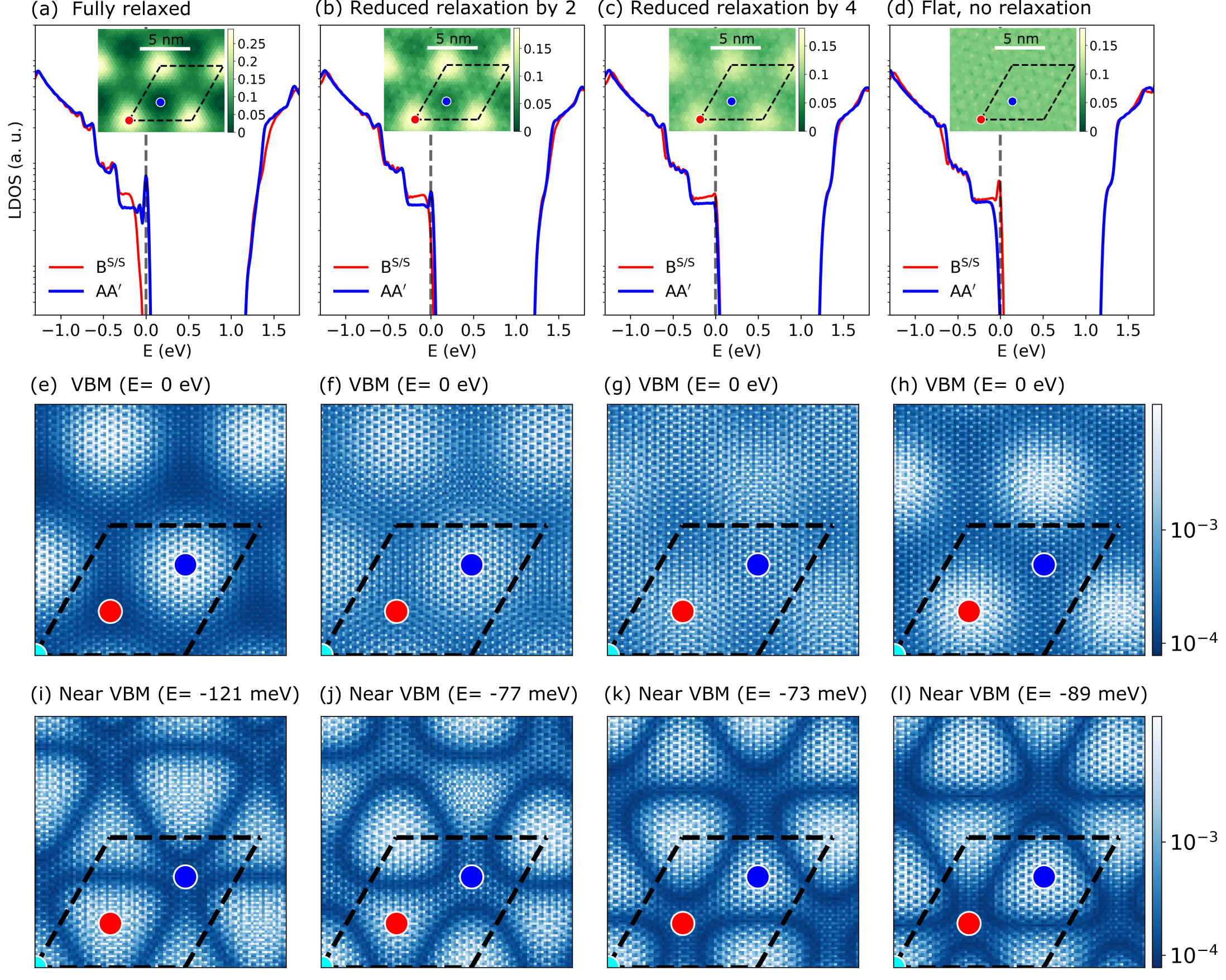}
    \caption{Calculated Local Density Of States (LDOS) from DFT for a free-standing twisted bilayer (a) including the full atomic relaxations, (b) with all atomic displacements reduced by a factor of 2 compared to the fully relaxed result, (c) with all atomic displacements reduced by a factor of 4 compared to the fully relaxed result and (d) without any atomic relaxations. The valence band maximum is set to 0. The insets show the calculated STM topographs. The associated color bars indicate variation in height in nm. (e)-(h): Squared wavefunctions of the highest valence band at the $\Gamma$-point of the moir\'e Brillouin zone for the atomic structures shown in the top panel. (i)-(l): Squared wavefunctions of a deeper lying valence band at the $\Gamma$-point of the moir\'e Brillouin zone for the atomic structures shown in the top panel.}
	\label{f4}
\end{figure}

To further understand the experimental findings, we calculated the local density of states (LDOS) at the AA' and $\mathrm{B^{S/S}}$ sites using DFT (see SI for more details). Figure 4(a) shows the result for a fully relaxed free-standing twisted bilayer (i.e. a bilayer in free space without a substrate). In the inset of Fig.~4(a), we also show the calculated STM topography. The LDOS in Figure~4(a) suggests that in this scenario the highest valence band is highly localized at the $\mathrm{AA}$' sites, contrary to the experimental findings of Figure 3. Interestingly, the highest valence band is localized on the $\mathrm{B^{S/S}}$ site when no atomic relaxations are included and the layers remain flat, see Fig. 4(d). This suggests that atomic relaxations in the experimental sample are not as strong as in a free-standing system. Possible origins of this discrepancy include (i) the influence of the substrate which is not included in our simulations, (ii) atomic motions due to thermal fluctuations, and (iii) extrinsic strains. To test this hypothesis, we also carry out DFT calculations for twisted bilayers in which the displacements of all atoms from their unrelaxed positions are reduced by a factor of 2 in Fig. 4(b) and by a factor of 4 in Fig. 4(c), compared to their values in the fully relaxed free-standing system. In Figure 4(h) we show the calculation in the absence of relaxation. In these systems, the LDOS in the AA' and the $\mathrm{B^{S/S}}$ regions are almost the same near the valence band onset in good agreement with the experimental findings. These results clearly demonstrate the high sensitivity of the electronic structure on atomic relaxations.

Figures 4(e)-(h) demonstrate that the wavefunction of the highest lying valence band does not correspond to the experimental measurement for the feature V$_1$ shown in the lower panel of Fig. 3: the wavefunctions do not form a honeycomb shape, but a triangular pattern. The only exception corresponds to the wavefunctions obtained when all displacements are reduced by a factor of 4, see Fig. 4(g). However, in this case, the dark regions of the honeycomb pattern are not associated with $\mathrm{B^{S/S}}$ regions as in the experiment. Instead, we have found that a deeper-lying valence band has a low intensity on the B$^{s/S}$ regions when relaxations are absent or strongly reduced, see Figs. 4(g) and (h). This state also exhibits a six-fold symmetry around these B$^{S/S}$ sites as does the honeycomb pattern observed in experiments. We, therefore, assign the experimental maps of the feature V$_1$ to this deeper-lying valence band.
Considering the thermal broadening in this experiment, it is not surprising that the contributions from a single narrow flat band would not be the only ones measured at a particular energy. 

In conclusion, we use scanning tunneling microscopy and spectroscopy to characterize the spatial distribution of the wave functions corresponding to flat bands in a  bilayer WS$_{2}$ sample twisted $\approx 3^{\circ}$ off the antiparallel alignment. To understand the localization at the different high-symmetry stacking regions we performed density-functional theory calculations which revealed that atomic relaxations play a critical role in determining the degree to which the wave functions associated with flat bands localize on the high-symmetry regions. We find that in our samples the atomic relaxations are much suppressed compared to calculations for the idealized system of a bilayer in vaccum. These findings provide insight into the spatial distributions of the wave functions associated with flat bands  believed to be responsible for the occurrence of correlated electron phenomena, demonstrating that atomic relaxations are a way to control the localization of the wave functions.

\begingroup
\renewcommand{\section}[2]{}


\subsection*{Methods}
The sample was prepared by the dry transfer method. Two monolayers of WS$_{2}$ (purchased from HQ Graphene) were stacked on each other using the tear-and-stack technique~\cite{boddison2019fabricating,boddison2019flattening}. Graphene (purchased from Graphenea) was used as a substrate to allow a sufficiently conductive contact for STM measurements. Both WS$_{2}$ and graphite crystals were mechanically exfoliated onto Si/SiO$_{2}$ substrates. Monolayer WS$_{2}$ was identified optically, and the thickness was confirmed with atomic force microscopy (AFM). Suitable graphite flakes were identified and were picked up with a polydimethylsiloxane (PDMS) stamp covered with Polypropylene carbonate (PPC) film on a glass slide. The graphite on the polymer stamp was then used to pick up monolayer WS$_{2}$. The two layers of WS$_{2}$ correspond to two halves of a large flake, allowing for precise rotational control. The PPC film was then cut and the 2D material stack was transferred onto a Si/SiO$_{2}$ substrate coated with Ti/Au (5 nm/150 nm). The sample was then annealed under ultra-high vacuum at 300°C for several hours. The surface was cleaned using an AFM tip~\cite{lindvall2012cleaning,rosenberger2018nano}. STM measurements were performed at pressures below $10^{-9}$ Torr, at room temperature. For STS we used bias modulation of 5.0 mV, 1.325 kHz and a typical set-point: $V_B = -1.3V$, $I_T = 150pA$.

\subsection*{Computational Details}
\subsection{Atomic reconstructions}
The twisted bilayer of WS$_{2}$ has been generated using the TWISTER package~\cite{Naiktwister2022}. The total number of atoms in the unit cell of $56.86^\circ$ twisted bilayer WS$_{2}$ is 1986. All the atomic relaxations are performed using classical interatomic potentials fitted to density functional theory calculations. The intralayer interactions within WS$_{2}$ are described using a Stillinger- Weber potential~\cite{Zhouhandbook2017}, and the interlayer interactions are captured using a Kolmogorov-Crespi potential~\cite{Naikkolmogorov2019}. All the simulations with classical interatomic potentials are performed using the LAMMPS package~\cite{Thompsonlammps2022,lammps}. We relax the atoms within a fixed simulation box with a force tolerance of 10$^{-6}$ eV/\AA\ for any atom along any direction. 

\subsection{Electronic structure calculations}
All the electronic structure calculations were performed using the SIESTA package~\cite{Solersiesta2002}. All the calculations included spin-orbit coupling~\cite{Seivaneon2006}. We used Troullier-Martins pseudopotential~\cite{Troullierefficient1991} and the local density approximation with the Perdew-
Zunger parametrization as the exchange-correlation functional~\cite{Perdewself1981}. We used a double-$\zeta$ plus polarization basis for the expansion of wavefunctions. For the electronic structure calculations we used the $\Gamma$ point in the moir\'{e} Brillouin zone to obtain the converged ground state charge density. A large vacuum spacing of $20$ \AA\ was used in the out-of-plane direction. 

\subsubsection{Density of states calculations}
To obtain well-converged results, the density of states calculations were performed on a fine $5\times 5\times 1$ k-grid in the moir\'{e} BZ (which corresponds to a $91\times91 \times 1$ k-grid of the monolayer Brillouin zone). We used a Gaussian with a width of 30 meV to represent the $\delta$-function in the density of states calculations. 

\subsubsection{Local density of states (LDOS) calculations}
To obtain the LDOS for selected stackings, we used a $5\times 5\times 1$ k-grid to sample the moir\'{e} Brillouin zone. Furthermore, we average the LDOS over a 9 \AA\ radius in-plane circular patch around the centre of the selected high-symmetry stackings. 

\subsubsection{STM topography calculations}
The calculations associated with the STM topography were performed only using the $\Gamma$ point in the moir\'{e} Brillouin zone. Furthermore, we used a Voronoi diagram to coarse-grain atomistic details in the topography. 

The codes developed for this work (selected LDOS and STM topography) will be made publicly available at \url{https://gitlab.com/_imaity_/Moire_toolkit_electronic}.

\subsection*{Acknowledgement}
AL-M, LM, RP, LA acknowledge funding from NSERC Discovery RGPIN-2022-05215, NRC Quantum Sensing Challenge, Ontario Early Researcher Award ER-16-218.     
This project received funding from the European Union’s Horizon 2020 research and innovation program under the Marie Skłodowska-Curie Grant agreement No. 101028468. This work used the ARCHER2 UK National Supercomputing Service (https://www.archer2.ac.uk) via our membership of the UK's HEC Materials Chemistry Consortium, which is funded by EPSRC (EP/R029431).

\clearpage
\subsection*{\large Supplementary information for: Moir\'{e} flat band in antiparallel twisted bilayer WS$_2$ relevaled by scanning tunneling microscopy
}
\beginsupplement

\subsection*{Section I: Different moir\'{e} patterns observed over the sample }

\begin{figure}
        \centering
	\includegraphics[width=0.6\columnwidth]{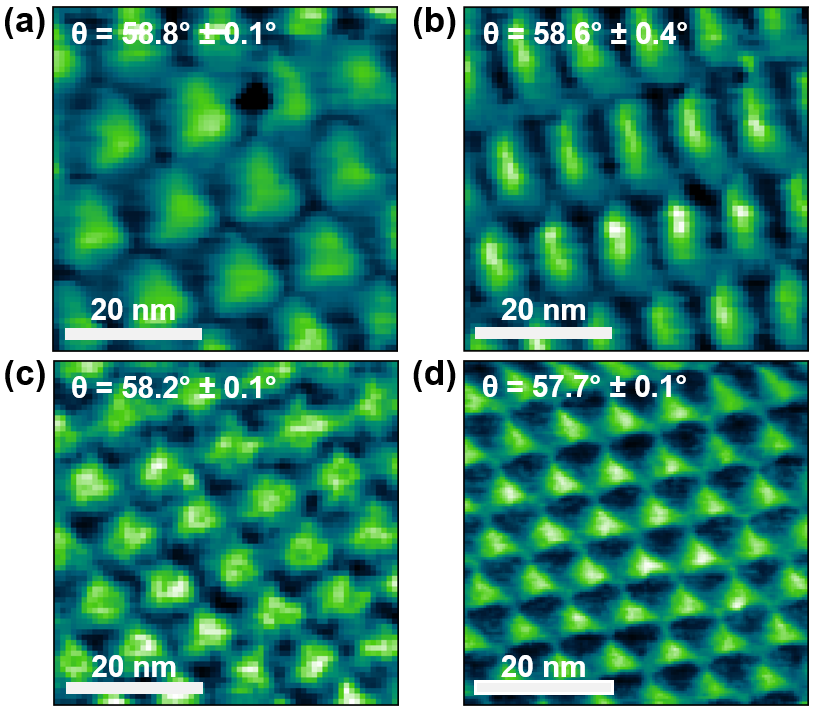}
    \caption{(a,b,c,d) STM topography with four different twist angles as indicated, corresponding to moir\'{e} wavelengths of 14 nm ± 1 nm, 13 nm ± 3 nm, 10.1 nm ± 0.7 nm, 7.9 nm ± 0.4 nm respectively. Data were taken at $V_{B} = -1.3 V$, $I_{t} = 100 pA$.}
	\label{f1}
\end{figure}

\clearpage
\subsection*{Section II: Spatial variation of the local density of states over moir\'{e} periods}

\begin{figure}
        \centering
	\includegraphics[width=0.6\columnwidth]{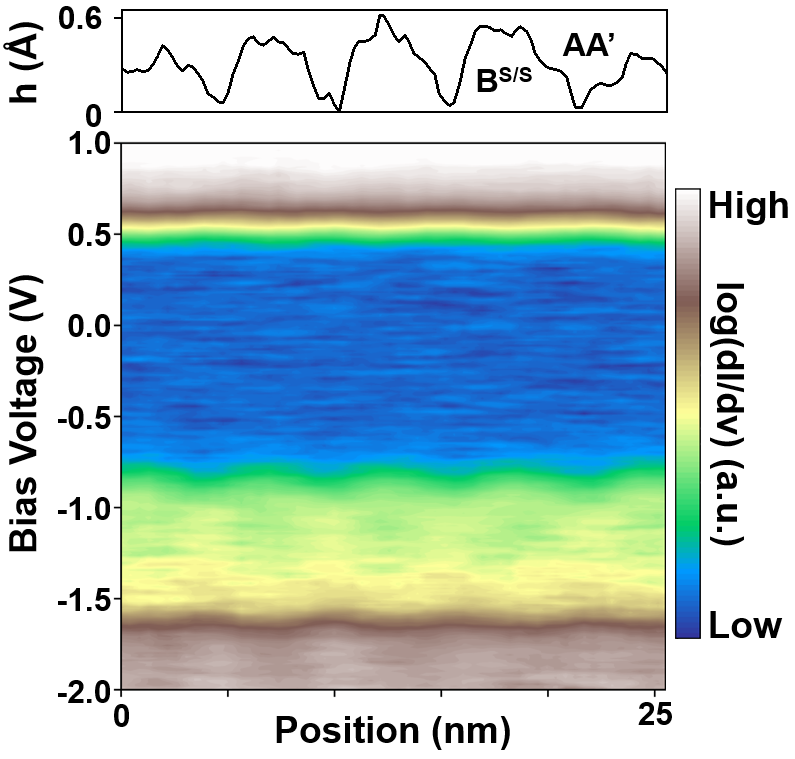}
    \caption{STS line profile crossing 4.5 moir\'{e} periods, demonstrating moir\'{e} modulation around $V_{1} = -0.9 V$ and $V_{2} = -1.6V$. The above figure indicates the height profile along the line indicated in the STM topography in Figure S3(a).}
	\label{f2}
\end{figure}
\clearpage
\subsection*{Section III: Spectroscopic gap calculated from the average spectra}

\begin{figure}
        \centering
	\includegraphics[width=0.6\columnwidth]{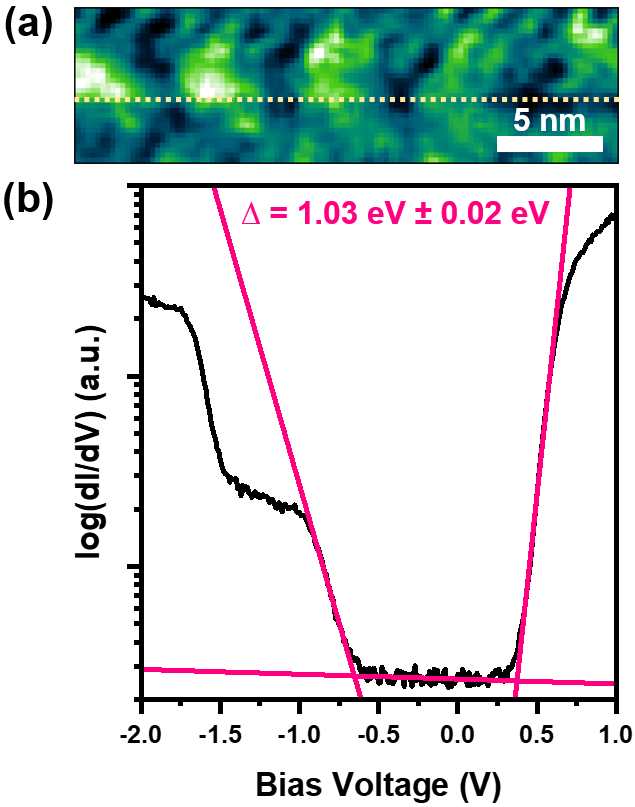}
    \caption{(a) Area in which Figure 2 (a) spectra where collected. Spectra in Figure 2 a) is the average of the curves collected along the indicated line. b) bang-gap estimation of scanning tunneling spectroscopy curve corresponding to Figure 2 (a). In a log plot, linear region of the valence and conduction band edges and the band gap are identified, and used to fit lines. The intersection of these lines correspond to the valence and conduction band edges as appropriate.}
	\label{f2}
\end{figure}

\end{document}